\documentclass[twocolumn]{jpsj2}

\def\runauthor{X. Zotos}
\title{Issues on the transport of one dimensional quantum systems}
\author{X. Zotos\thanks{e-mail: zotos@physics.uoc.gr}}
\inst{Department of Physics, University of Crete \\ and Foundation
for Research and Technology-Hellas, \\P. O. Box 2208, 71003 Heraklion,
Crete, Greece}
\recdate{\today}
\abst{In this paper, we will critically discuss recent theoretical and 
experimental developments on the transport of one dimensional 
quantum systems. In particular, we will focus on open issues and 
controversial results related to the finite temperature dynamics 
of integrable quantum models. These singular systems are commonly used in 
the description of quasi-one dimensional materials. They have recently 
attracted a lot of interest in connection to the interpretation 
of experiments, specially after the discovery of 
unusual thermal conductivity in magnetic compounds .}

\kword{quantum transport, integrability, quasi-one dimensional materials}

\begin{document}
\maketitle

\section{An old story}
In 1834 J. Scott-Russel observed\cite{russel} a ``singular and 
beautiful phenomenon" in the form of ``a large solitary elevation...which 
continued his course along the channel apparently without change of form or 
diminution of speed...". This phenomenon recently resurfaced in a quantum 
version remaining as exciting as well, as we will discuss extensively 
in this paper, singular.

Its mathematical description was formulated some years later by Korteweg 
and de Vries\cite{kdv} who proposed a one dimensional classical 
nonlinear wave equation to describe it. This 
equation (so called KdV) has nonlinear solutions in the form of 
localized, nondispersing 
waves. But it was only in the 1960's, with the development of large scale 
computer simulations\cite{zk} on dynamical systems, that it was realized 
that not only a single 
solitary wave solution was stable but that a set of ``solitary waves" entering 
into collision they emerge after the collision with unchanged velocities and 
shapes and only suffering a phase shift. 

Soon after this discovery, an impressive mathematical theory 
was proposed 
- the {\it Inverse Scattering Method} (ISM) - that solves the problem of 
time evolution 
of an initial configuration in an classical nonlinear 
but {\it integrable system} by a series of linear 
operations. In other words, this procedure is the analogue of the Fourier 
Transform for linear systems\cite{ggkm}.
Within this theory it was also discovered that integrable systems are 
characterized by a macroscopic number of conservation laws 
that we believe are the key to their singular transport properties.

At this point we should mention a distinction between systems with,  
(i) ``topological" solitons (the stability of which is guaranteed by   
topological constraints) that in general do not exhibit particular 
transport properties and, (ii) ``mathematical" solitons, 
the stability of which is guaranteed by the very specific structure of 
the Hamiltonian describing them\cite{soliton}. 
A possible source of confusion are systems as the  
sine-Gordon theory (classical and quantum) that are integrable but they 
also possess topological excitations. 

The paradigm of {\it classical integrable} systems finds now-days applications
in all branches of science and engineering. We will just mention the 
example of nonlinear optics where the coding and transmission of 
information is by optical solitons in 
fibers described by the nonlinear Schr\"odinger equation\cite{optsol}, 
a classical integrable model.
A key issue for applications, and as we will discuss below also for 
quantum ones,
is the robustness of dissipationless transport to perturbations that 
destroy the integrability of the underlying system.

\section{The quantum world}
In parallel to these developments, the first {\it quantum integrable} 
system was 
discovered by H. Bethe in 1931\cite{bethe} who proposed an exact form for the 
eigenfunctions and eigenvalues for the prototype spin-1/2 one dimensional 
isotropic Heisenberg model. Over the years, a particular technique was 
developed, the {\it Bethe ansatz} (BA) method, that allows to extract 
physical information 
from the rather complex eigenfunctions for a large class of quantum 
integrable models; for instance the Heisenberg, Hubbard, supersymmetric 
t-J model\cite{korepin,takahashi}, to cite a few examples of interest 
in the field of Condensed Matter Physics.
At the moment we can say that the thermodynamic properties of these models 
can analytically be studied and compared to experiment but that the finite 
temperature dynamic correlations, most relevant e.g. to transport 
experiments, are not easy to handle.

In a key development in 1979, by L. Fadeev and collaborators, it was realized 
that the Bethe ansatz method is nothing else but a quantum version of the 
ISM and thus there is a close analogy between classical 
and quantum integrable systems\cite{korepin}. In particular, they have 
developed a similar method, 
the {\it Quantum Inverse Scattering Method} (QISM).  
Within this formalism, it can be shown that 
the quantum systems are also characterized by a macroscopic number of 
conservation laws, that can be systematically constructed.

A fundamental issue emerging from this analogy, and being a driving force 
in this field, is the extent to which the physics and 
applications of classical integrable systems can be carried over to the 
the quantum ones, for instance in the context of unconventional transport 
properties of (quasi-) one dimensional materials described by quantum 
integrable Hamiltonians.

\section{Framework}

Most of the studies on transport properties of one dimensional quantum 
many body systems over the last few years were within the 
linear response theory 
(Kubo formalism\cite{kubo}). One fundamental quantity that attracted particular 
attention is the {\it Drude weight} $D$, defined as the weight of the 
zero-frequency contribution to the real part of the conductivity,  
\begin{displaymath}
\sigma'(\omega)=2\pi D\delta(\omega)+\sigma_{reg}(\omega > 0)
\end{displaymath}

\noindent
or equivalently as the pre-factor of the  reactive 
response $\sigma''(\omega)=D/\omega|_{\omega \rightarrow 0}$ at low 
frequencies. 
As it follows from this definition, a finite $D$ implies a ballistic 
response. Note also for further use, that the d.c. conductivity is given by the 
$\omega\rightarrow 0$ limit of the regular part, 
$\sigma_{dc}=\sigma_{reg}(\omega\rightarrow 0$).

The Drude weight was first proposed by W. Kohn in 1964 as a criterion
of metallic or insulating behavior in the context of the Mott-Hubbard 
transition\cite{kohn}. In particular, it was related to the response 
of the ground state energy, of a system with periodic boundary 
conditions forming a ring, to a fictitious flux $\phi$ piercing the ring.

Recently, the Kohn expression for the Drude weight was generalized at   
finite temperature\cite{czp} $T$. It is currently 
extensively studied in order to characterize the conduction of a 
system as ideal (ballistic - dissipationless) or diffusive.
In an alternative, perhaps more physical, interpretation it can be shown 
that it is essentially proportional to the long time asymptotic 
value of the (e.g. charge, spin or energy) current-current 
correlations\cite{znp}. 
The behavior of the Drude weight, although a seemingly simple quantity,  
still defies reliable characterization in several integrable models that 
have been studied so far. This is because it is not a 
thermodynamic quantity and thus difficult to analyze within known 
analytical methods.

Similarly, a thermal Drude weight $D_{th}$ can be defined that characterizes 
the thermal conductivity $\kappa(\omega)$ and it is proportional to the long 
time asymptotic of the energy current-energy current dynamic correlations.

The finite frequency behavior of the conductivity is still more involved 
and most known results are based on low energy effective theories of the 
Hamiltonians of interest or numerical simulations; for instance, variants of 
Exact Diagonalization (ED) via the Lanczos procedure (for a recent 
technique see the Microcanonical Lanczos Method\cite{lpskz} and references 
therein) and the Quantum Monte Carlo method\cite{gros1,kirchner}. 

The numerical results are however limited to rather small size 
lattices considering the exponentially fast increasing number of states 
in quantum many body systems. 
This implies, on the one hand, limitations on the information 
that can be extracted on the long time, equivalently low frequency - $\omega$, 
behavior of the conductivities. 
On the other hand, it also implies limitations on the smallest 
wave-vector - $q$ correlations, of the order of $1/L$, that can be evaluated. 
Thus it is difficult to explore the $q,\omega\rightarrow 0$ region, 
relevant to the hydrodynamic regime. 
Notice for instance, that any finite size system shows 
a nonzero Drude weight (as there is always a part of coherent transport 
in such a system), that however scales to zero in the thermodynamic limit
if the system shows normal transport.
From this observation we can also conclude that in numerical simulations it is 
more advantageous and reliable to study the high temperature limit where 
we expect a relevant ``mean free path" to be shorter, hopefully 
less, than the size of the lattice.

We should also mention a recent, promising development, 
where by the Density Matrix 
Re-normalization Group method\cite{dmrg} (that has been proven so successful
in the study of ground state properties) the time evolution of an initial 
state can evaluated on fairly large size lattices. 
Further development of this method might give access at the finite temperature, 
out-of equilibrium dynamics of one dimensional quantum lattice systems.

So we conclude that it remains crucial, in this field of 
strongly correlated systems, the development of reliable numerical simulation 
techniques for the study of finite temperature dynamic correlations. 
As they are related to transport are the most interesting for making 
contact with experiment.

\section{A conjecture}

\begin{figure}
\begin{center}
\includegraphics[width=8cm]{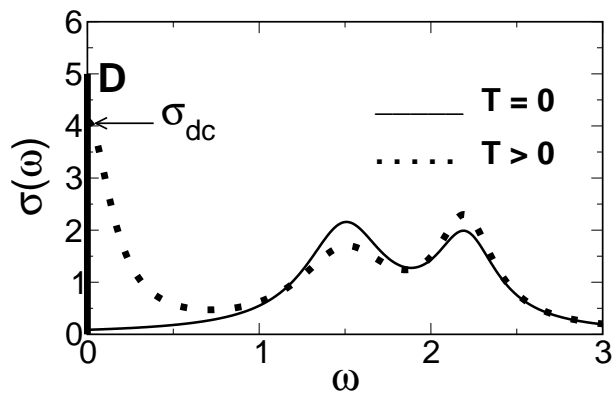}
\end{center}
\caption{Illustration of typical behavior of the conductivity in a 
nonintegrable - generic - system showing normal transport.}
\label{f1}
\end{figure}

A few years ago\cite{saito,mccoy,czp}, it was conjectured that integrable 
quantum many body systems show fundamentally different finite temperature 
transport properties than the - generic - nonintegrable ones. 
According to the common scenario, illustrated in Fig. \ref{f1}, 
a normal metallic system, with no disorder, at zero temperature, shows a finite 
Drude weight $D$ and thus a diverging d.c. conductivity.
As the temperature rises, due to Umklapp scattering, the zero-frequency 
Drude $\delta-$function broadens to a 
peak of width of order $1/\tau$ where $\tau$ is a characteristic scattering 
time. Thus a finite d.c. conductivity $\sigma_{dc}$ is obtained that 
decreases with increasing temperature and decreasing $\tau$. 

\begin{figure}
\begin{center}
\includegraphics[width=8cm]{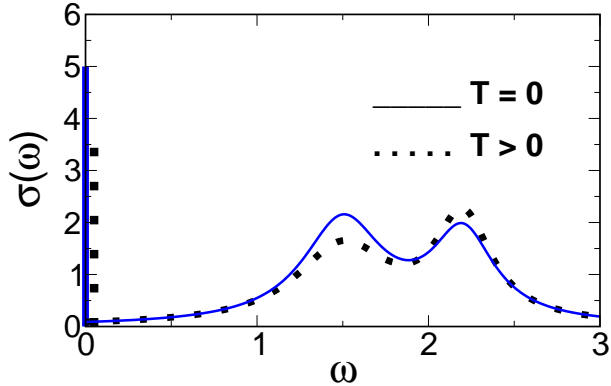}
\end{center}
\caption{Illustration of typical behavior of the conductivity in an 
integrable system showing ballistic transport at all temperatures.}
\label{f2}
\end{figure}

In contrast, as it is illustrated in Fig. \ref{f2}, in an integrable system 
this broadening  with temperature does not occur. 
The Drude weight remains finite at all temperatures, but with a 
nontrivial temperature dependence. Thus the system remains an ideal 
conductor, exhibiting ballistic transport at all $T$. 
Note that we are concerned with the conductivity of bulk systems. 
Finite systems with boundaries would show a vanishing $D$ 
and a d.c. conductivity proportional to the scattering 
by the boundaries. 

Two main issues, arising in relation to this conjecture, are debated 
at the moment.  

First, it was proposed that even nonintegrable systems  
might also show ballistic transport at finite $T$\cite{gros2,brenig,fk2};
evidence so far indicates that it is probably not the case.
To appreciate the difficulty of this question, note that despite 
extensive studies over the years in classical nonlinear one 
dimensional systems, it is still debated whether 
the thermal conductivity might be diverging\cite{livi} even in 
nonintegrable ones. In particular, in several nonintegrable models,  
it was found that although the Drude 
weight (long time asymptotic of current correlations) vanishes, the 
integral over time of the dynamic correlation 
(that gives the d.c. conductivity) diverges with system size. 
Furthermore, the simulations were carried over on systems with thousands 
of lattice sites, that indicates the subtlety of the scaling procedure.

Second, it is unclear whether an integrable model in the gapped phase 
(insulating) at zero temperature (e.g. half-filled 
Hubbard, easy-axis $S=1/2$ Heisenberg, nonlinear $\sigma$ model)
turns to an ideal conductor or a normal metal at finite temperatures
\cite{zp,fk,kirchner,carmelo,pszl,sachdev2,fujimoto,konik}.

Finally, we want to remark that although most of the studies focused on 
the behavior of the Drude weight, it still possible that $D$ might vanish 
but the low frequency behavior of the conductivity will turn out to be 
unconventional, e.g. power law, implying a diverging d.c. conductivity or 
of a non-diffusive form.

\section{Three Heisenberg models}

Spin models are prototype quantum many body systems that offer the 
possibility to test the above conjecture. Their study is also 
relevant in the interpretation of experiments in quasi-one dimensional 
magnetic compounds.

In this section we will discuss the transport behavior of three one 
dimensional spin models, (i) one integrable, the $S=1/2$ Heisenberg 
chain and (ii) two nonintegrable ones, the $S=1$ and $S=1/2$ 2-leg ladder 
systems. We will focus on open issues and discuss contradictions in the 
results obtained by different approaches.

\subsection{S=1/2}

The Hamiltonian is given by the generic form,
\begin{equation}
H=J\sum_{i=1}^L (S_i^x S_{i+1}^x +
S_i^y S_{i+1}^y + \Delta S_i^z S_{i+1}^z) ,
\label{heis}
\end{equation}

\noindent
where $S_i^{\alpha}$ are components of $S=1/2$ operators at site $i$ and 
we consider the, $J>0$, anti-ferromagnetic regime.
For $|\Delta| < 1$, the ``easy-plane" case, the spectrum is gapless, while
for $\Delta > 1$, the ``easy-axis" case, it is gapped.
By a Jordan-Wigner transformation, this spin model is equivalent to a 
spinless fermion system with hopping $-t$ and nearest-neighbor 
interaction $V$; the correspondence of parameters is $\Delta/J=V/2t$.
The model is integrable and its Bethe ansatz algebraic structure\cite{korepin} 
and thermodynamic properties have been extensively
studied\cite{takahashi}. Regarding the finite temperature dynamics, 
the picture on the spin conductivity is still debated while the thermal 
one is rather clear.

First, it is easy to show that indeed this integrable system  
can exhibit ideal (ballistic) transport at all temperatures by using the fact 
that it is characterized by nontrivial conservation laws. 
In particular, it was shown\cite{znp}, using an inequality proposed by 
Mazur and Suzuki\cite{mazur}, that $D$ at high temperatures, 
$\beta=1/\kappa_B T \rightarrow 0$, 
is bounded by the ``overlap" of the 
spin current operator $j^z$ with at least the first nontrivial 
conservation law $Q_3$ as ($m=<S_l^z>$),

\begin{equation}
\begin{split}
D(T)&\geq \frac{\beta}{2 L}\frac{<j^z Q_3>^2}{<Q_3^2>}\\
&= \frac{1}{2k_B T} \frac{8 \Delta^2 m^2 (1/4-m^2)}
{{1+8\Delta^2(1/4+m^2)}}.
\end{split}
\label{dheis}
\end{equation}

This idea, although very powerful in proving that such a state - ideal 
conductor at all temperatures - exists, it provides at the moment a 
rather incomplete picture of the behavior of $D$. Indeed, we see from 
Eq. (\ref{dheis}) that the r.h.s. vanishes at $\Delta=0$, the point that 
corresponds to the $XY$ model, 
where we know that the spin current commutes with $H$ and thus $D$ is finite. 
Also, at magnetization 
$m=0$, the r.h.s. also vanishes according to Eq. (\ref{dheis}) 
while independent BA\cite{xz1,kluemper} 
and numerical evaluations\cite{zp,nma,gros1,brenig}  
show that it is finite in the $\Delta < 1$ regime. 
The same conclusion holds even including all 
conservation laws derived from the algebraic structure of this integrable 
model\cite{korepin} as it can be seen by invoking their ``electron-hole" 
symmetry\cite{znp}. 

\begin{figure}
\begin{center}
\includegraphics[width=8cm]{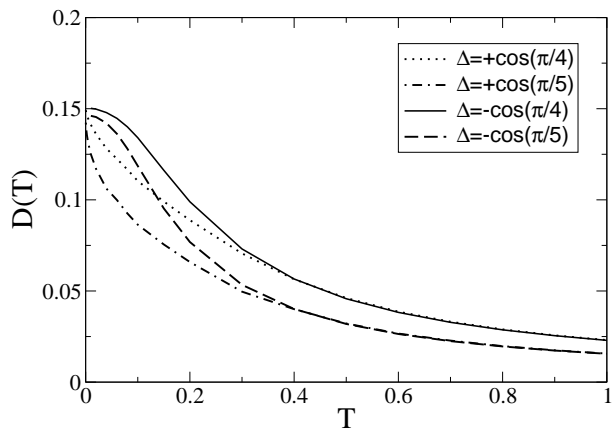}
\end{center}
\caption{Temperature dependence of the Drude weight\cite{xz1} for 
the $S=1/2$ Heisenberg chain model and 
different values of the anisotropy parameter $\Delta$. }
\label{f3}
\end{figure}

Regarding the BA calculation\cite{xz1} at $m=0$, it is based 
on the assumption of a ``rigid string" picture for the excitations 
proposed by Fujimoto and 
Kawakami\cite{fk} in the context of the Hubbard model; notice that it can 
be done at the special values of $\Delta=\cos(\pi/\nu),~~~\nu=$integer. 
It predicted a nonzero $D$ for $|\Delta| < 1$ with a power law behavior 
at low temperatures as it can be seen in Fig. \ref{f3}. 
In particular: 

\begin{itemize}
\item at zero magnetization, in the easy plane anti-ferromagnetic regime
($0<\Delta<1$), the Drude weight decreases at low
temperatures as $D(T)=D_0-{\rm const.}T^{\alpha},
~~~\alpha=2/(\nu-1),~~~\Delta=\cos(\pi/\nu)$;

\item in the ferromagnetic regime, $-1<\Delta < 0$, $D(T)$ decreases
quadratically with temperature (as in a noninteracting, XY-system);

\item the same low temperature quadratic behavior is true at any finite
magnetization;

\item for $\beta\rightarrow 0$, $D(T)=\beta C_{jj}$
and it can be shown that $D(-\Delta)=D(\Delta)$
by applying a unitary transformation;

\item a closed expression for $C_{jj}$ has be obtained analytically 
\cite{kluemper},
$C_{jj}=(\pi/\nu-0.5 \sin(2 \pi/\nu))/(16\pi/\nu)$ for $|\Delta|<1$, 
while $C_{jj}=0$ for $\Delta>1$;

\item at the isotropic anti-ferromagnetic point ($\Delta=1$), $D(T)$
seems to vanish, implying non ballistic transport at all finite
temperatures.
\end{itemize}

The above low temperature behavior for $|\Delta|<1$ is challenged by an 
alternative BA ``spinon" approach\cite{kluemper}, QMC simulations\cite{gros1}
and an effective field theory approach\cite{fk2}, although all agree 
on a finite $D$ in this regime.
On this disagreement, we can comment that, in the QMC simulations 
the temperature is very 
low (so as to obtain reliable real time data from imaginary time ones)  
of the order of the level spacing and thus not necessarily in the 
bulk regime - level spacing much less than  $T$.
Second, in effective field theories the dispersion relation is linearized and 
it is not clear whether curvature effects might affect the temperature 
behavior of $D$; note that, for free fermions on a lattice, $D$ decreases 
as $T^2$ at low $T$ while with a linearized dispersion is constant.

Regarding low energy effective theories, the canonical approach 
to one dimensional quantum systems, the Luttinger liquid 
theory, was applied to transport\cite{giamarchi}. In the context of 
this approach, it was recently stressed the importance of, 
(i) conservation laws\cite{rosch} in altering the behavior of 
transport - from diffusive to ballistic - 
and, (ii) correctly accounting perturbations irrelevant to  
thermodynamics but crucial for transport\cite{shimshoni}.
We should emphasize that it remains an outstanding problem, first, 
to match the descriptions in terms of 
the original (non)integrable Hamiltonian and its effective field 
theory, second, to account for curvature effects and, third, to trace the 
fate of conservation laws - present at all energies in BA systems - 
to low energies.

\begin{figure}
\begin{center}
\includegraphics[width=8cm]{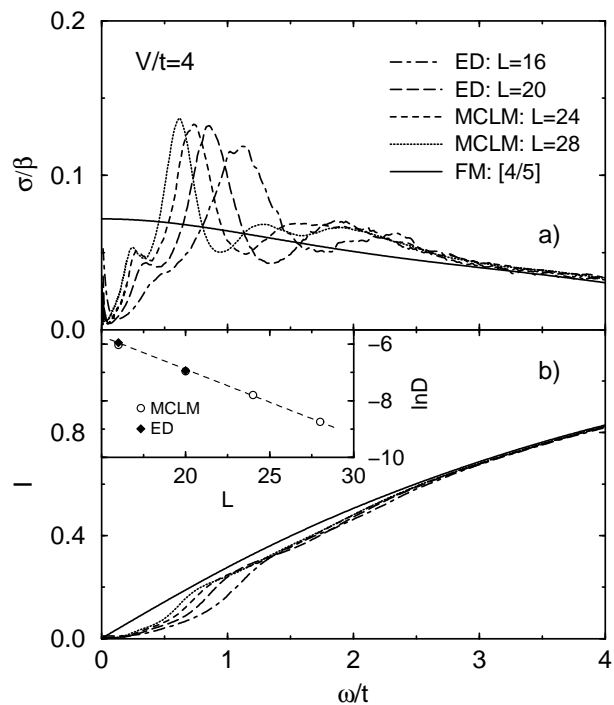}
\end{center}
\caption{Low frequency conductivity\cite{pszl} 
in the gapped regime of the $S=1/2$ 
Heisenberg model, $\Delta/J=V/2t$; $I$ is the integrated in frequency 
conductivity.}
\label{f4}
\end{figure}

For $\Delta > 1$ the Heisenberg model is gapped and thus $D=0$ at zero 
temperature. Numerical evidence in the $\beta\rightarrow 0$ limit\cite{pszl} 
(shown in Fig. \ref{f4}) and Bethe ansatz analytical 
results\cite{xz1,kluemper}  
suggest that $D$ remains zero at all temperatures. 
The data follow from Exact Diagonalization calculations for the smallest 
systems and 
the Microcanonical Lanczos method (MCLM) for the largest ones, while FM 
denotes a frequency moment analysis. 
Note the crucial role of the MCLM technique in extracting sufficient 
information from the conductivity spectra. 
Very large finite size effects are observed in the low frequency behavior 
of the conductivity that scale as $1/L$ and the significance of which is not 
clear at the moment. 
It is interesting to examine whether they signify 
an unconventional low frequency-small wave-vector behavior that is not 
of the standard diffusive form. 
They also render particularly difficult 
the analysis of the bulk system conductivity and the extraction, 
if there is one, of a value for the d.c. conductivity. 
Also note that, one cannot exclude the convergence of the finite frequency 
peak to a Drude $\delta-$function in the thermodynamic limit, although 
such a scenario is rather unlikely.
We should remind that a semi-classical theory for gapped systems 
by Sachdev and collaborators predicts a normal diffusive 
behavior\cite{sachdev2}.

Finally, we will conclude with some open questions related to the issue of 
conservation laws and transport. From the Mazur inequality one could argue 
that the presence of only one conserved quantity is enough to guarantee 
a finite $D$ and that integrability is not a necessary condition. 
However, there are arguments that, at least 
within the BA systems, the presence of one conservation law guarantees 
the existence of a macroscopic number. Notice that for the 
$S=1/2$ Heisenberg\cite{grab} and the Hubbard model\cite{ladder} 
the conservation laws can be constructed recursively using a ``boost" operator.
Conversely, from the case $m=0$, it is not clear that the conservation 
laws, in the context of the Mazur-Suzuki (in)equality, exhausts the Drude 
weight or even that the presence of 
at least one is necessary for a finite $D$. 
Finally, on this problem, we can also 
argue that if the BA analysis\cite{xz1} turns out to be valid and $D$ 
shows a non-analytic behavior at $\Delta=1$, then one, or a finite 
number of conservation laws within the Mazur inequality cannot 
reproduce this behavior.

Concerning energy transport the situation is far more clear and simple. 
It was early on 
realized\cite{nie,huber} that the energy current in the $S=1/2$ Heisenberg 
model commutes 
with the Hamiltonian, at it happens to coincide with the so-called $Q_3$ 
conservation law in the BA analysis. Thus it follows that the energy current 
correlations do not decay in time and thus the thermal conductivity diverges. 
The ``thermal Drude weight" has been exactly evaluated using Bethe ansatz 
techniques\cite{kluemper2} for all values of the anisotropy $\Delta$. 

Based on the same idea, of a commuting energy current with the Hamiltonian, 
we finally mention a magneto-thermal effect that has recently 
been proposed\cite{louis}. In this phenomenon, a spin current would 
flow in the presence of a temperature gradient and a magnetic field, 
a phenomenon analogous to the thermoelectric effect for electronic 
systems. The study, using ED and QMC methods, found 
a diverging transport coefficient resulting from the conservation of the 
energy current.

\subsection{S=1}

A natural question coming up is, what about the transport behavior of the 
same Heisenberg model Eq. (\ref{heis}) but for higher spin, e.g. $S=1$, that is 
not an integrable model. The $S=1$ case is known to be a gapped 
system\cite{haldane} and thus an insulator at $T=0$ with $D=0$. 

Its low energy physics is described by the 
quantum nonlinear $\sigma$-model (qNL$\sigma$M). 
A semiclassical treatment of this effective theory by Sachdev and 
collaborators\cite{sachdev}, that essentially maps this system to  
a classical gas of different species impenetrable particles, concluded
to diffusive dynamics. Note, that the classical model is also integrable 
and can be analytically solved. 
On the other hand, as the nonlinear $\sigma$ model is also 
an integrable field theory, the Bethe ansatz method has been 
applied\cite{fujimoto,konik} but there it was concluded that the Drude 
weight is finite and thus the transport ballistic.
This discrepancy raises the question whether, within the qNL$\sigma$M model 
description, subtle quantum effects due the 
finite tunneling of the quasi-particles alter the transport behavior from 
diffusive to ballistic.

\begin{figure}
\begin{center}
\includegraphics[width=8cm]{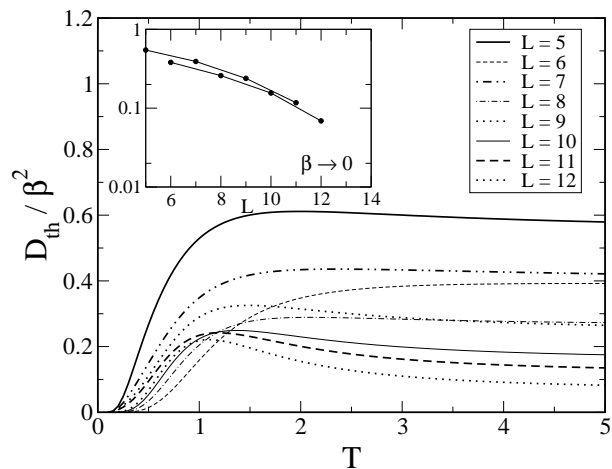}
\end{center}
\caption{Scaling of the thermal Drude weight $D_{th}$ for the $S=1$ Heisenberg 
spin chain model\cite{kz}.}
\label{f5}
\end{figure}

\begin{figure}
\begin{center}
\includegraphics[width=8cm]{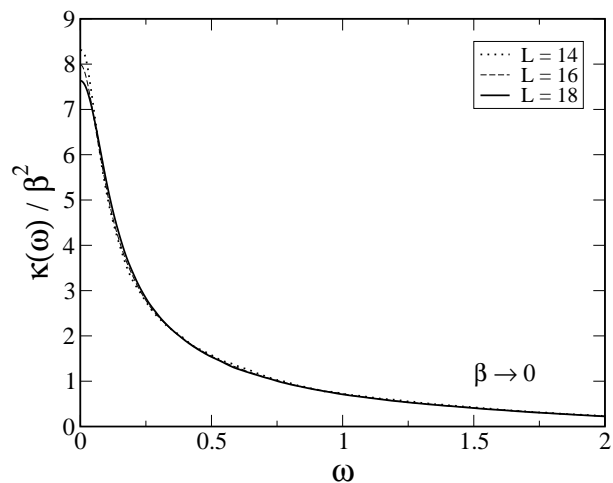}
\end{center}
\caption{Thermal conductivity of the $S=1$ Heisenberg isotropic spin chain 
model in the high temperature limit\cite{kz}}
\label{f6}
\end{figure}

\begin{figure}
\begin{center}
\includegraphics[width=8cm]{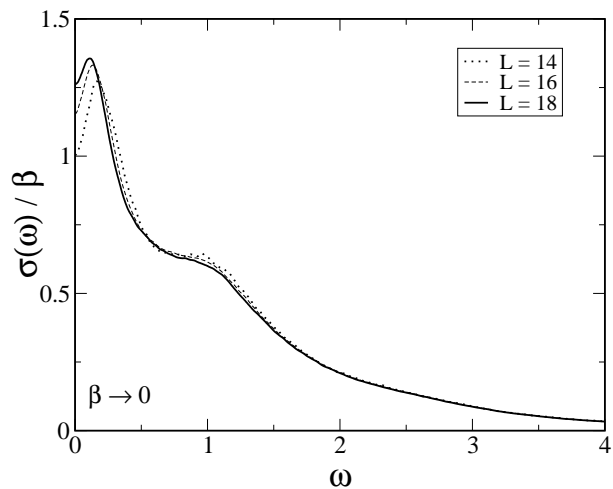}
\end{center}
\caption{Spin conductivity of the $S=1$ Heisenberg isotropic spin chain 
model in the high temperature limit\cite{kz}.}
\label{f7}
\end{figure}

Motivated by this problem and the theoretical and experimental interest 
for the dynamics of $S=1$ compounds, a numerical simulation in the high 
temperature limit\cite{kz} was performed. It 
indicates that both the spin and energy conductivities have 
finite d.c. values. 
The frequency dependence of the data however do not seem to fit a standard 
Lorenzian/diffusive form.
The results for $\kappa({\omega})$ and $\sigma({\omega})$ 
are shown in Fig. \ref{f6} and \ref{f7} respectively. 
It became possible to obtain them  
for the largest lattices after the development of the MCLM that allows to 
reach as low frequencies as possible and minimize statistical noise 
due to finite size effects.

But first, the value of the Drude weight in the thermodynamic limit 
must be deduced as shown in Fig. \ref{f5}. 
On this point, we should mention that finite temperature data  
might create some confusion because, at low 
temperatures, for systems with an even number of sites, 
the Drude weight seems to scale upwards with increasing 
system size while at high temperatures it seems to scale downwards, to zero.
This observation could lead to the suggestion that the transport is ballistic 
in this nonintegrable system.
We think however, that the low temperature data should be considered with 
care as the crossing point in temperature where the scaling direction reverses 
moves to lower temperatures with increasing system size. Thus   
the region of temperature where the scaling is upwards will shrink to zero 
in the thermodynamic limit. Also note that in lattices with an odd number of 
sites $D_{th}$ scales downwards to zero at all $T$.

The conclusion that in the $S=1$ isotropic Heisenberg model the d.c. 
conductivities are finite means that either the assumptions in the BA 
analysis of the nonlinear $\sigma$ model are not warranted or 
that in the truncation from the full $S=1$ model to the effective one, 
relevant for transport terms are omitted that turn the diffusive transport 
to ballistic.

We should also emphasize the drastic difference, as seen by comparing 
Fig. \ref{f4} and Fig. \ref{f7}, in the finite size dependence 
of the low frequency conductivity between the integrable $S=1/2$  
Heisenberg model for $\Delta>1$ and the $S=1$ model, both gapped systems. 
In the $S=1$ system, finite size effects are limited to exponentially 
small, with system size $L$, frequencies while 
in the $S=1/2$ to a $1/L$ frequency range. 
Although both might converge to a finite d.c. conductivity, the 
significance of this difference in scaling is an open issue at the moment.

\subsection{S=1/2 2-leg ladder}

The next question that must be addressed is the robustness of ideal transport
of an integrable system to perturbations breaking integrability.
Namely, whether the presence of an integrable point in interaction parameter 
space is signaled by an enhanced conductivity and this over a finite region 
and not an infinitesimal one, e.g. one that shrinks to zero as the system size 
increases to the bulk limit. 

In this context, frustrated spin chain and ladder models have been 
studied\cite{gros2,brenig} by numerical diagonalization techniques with 
conclusions that are still debated, although it seems that the 
integrability-transport conjecture  seems to hold. However, we should keep 
in mind that the conclusions are drawn from numerical studies on 
rather small 
size lattices and thus extreme care is needed in drawing definite 
conclusions for the bulk system; no exact results are known so far.

A system on which the question of robustness can be studied is 
the $S=1/2$ 2-leg ladder 
model described by the Hamiltonian,

\begin{equation}
H=J\sum_{l=1,L}  ({\bf S}_{1,l+1} \cdot {\bf S}_{1,l} +
{\bf S}_{2,l+1} \cdot {\bf S}_{2,l})
+\frac{J_{\perp}}{J}  {\bf S}_{1,l} \cdot {\bf S}_{2,l} ,
\label{ladder}
\end{equation}

\noindent
where $S_i^{\alpha}$ are components of $S=1/2$ operators at site $i$ and 
we consider the, $J>0$, anti-ferromagnetic regime.
The interchain coupling $J_{\perp}$ controls the perturbation 
breaking the integrability of the separate Heisenberg chains.
Indeed, for $J_{\perp} \rightarrow 0$ the two chains decouple and the 
thermal conductivity of the ladder is the sum of the conductivities 
of each chain, that diverge as the energy current of the Heisenberg 
chain model commutes with the Hamiltonian.

This model is also of experimental interest as the finite temperature 
transport of several compounds described by this Hamiltonian are recently 
investigated.
The low energy physics of this gapped system is also described by 
the nonlinear $\sigma$ model and extensive studies\cite{sachdev2,orignac} 
based on this approach have been performed. In particular, the analysis by 
Sachdev and collaborators concluded to a fairly complete picture of 
diffusive dynamics.

\begin{figure}
\begin{center}
\includegraphics[width=8cm]{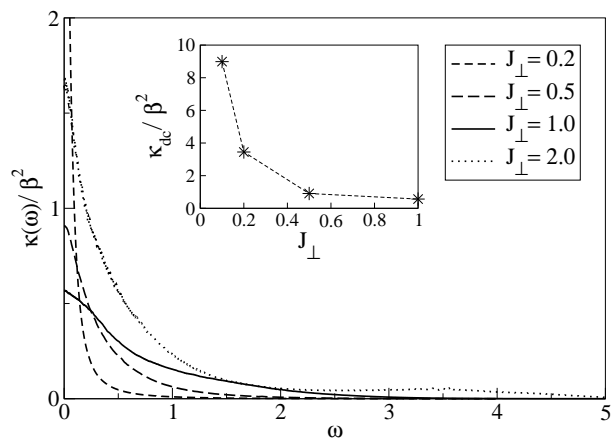}
\end{center}
\caption{Thermal conductivity of the $S=1/2$ 2-leg ladder 
model in the high temperature limit\cite{xz2} for $L=16$.}
\label{f8}
\end{figure}

A recent numerical simulation study in the high temperature limit\cite{xz2}, 
again using the MCLM, also indicates a finite d.c. thermal conductivity as 
shown in Fig. \ref{f8}. The frequency dependence, as in the $S=1$ model, 
does not seem to conform to a diffusive form. Furthermore, as shown in the 
inset, the d.c. thermal conductivity scales as 
$(J/J_{\perp})^2$, diverging as it should in the limit of decoupled chains. 
This result is interesting because, first, it indicates that the influence 
of an integrable point ($J_{\perp}=0$) extends to a finite, of order one, 
region in interaction parameter space and, second, the quadratic 
dependence on the 
interaction is similar to that expected from a perturbative result. 
Notice however that the perturbation now is around not an independent 
particle system, but around a fully interacting, albeit integrable, one.

Thus the conclusion of a finite d.c. conductivity at high temperatures 
does not contradict the semiclassical treatment of the qNL$\sigma$M at low 
temperatures, but the same problematic 
as the one for the $S=1$ model applies in this case. 

\section{A new mode of transport}

Quasi-one dimensional materials have been synthesized and experimentally 
studied since the 60's. Recently there has been renewed interest in their 
transport properties as new materials where discovered - some in the 
class of cuprates - with very low disorder, very weak interchain coupling 
and extraordinary transport properties.

In relation to the above theoretical developments we should mention the  
NMR experiments by Takigawa and Thurber\cite{takigawa,thurber} on $S=1/2$  
compounds (e.g. $Sr_2CuO_3$) that show diffusive spin dynamics but with 
a diffusion constant orders of magnitude larger than that expected from a 
classical moment analysis. Similarly, the spin dynamics of the $S=1$ 
compound $AgVP_2S_6$ has been investigated, indicating diffusive behavior in 
this case\cite{takigawa1}, at least at elevated temperatures.

The phenomenon that attracted most interest recently\cite{k1,k2,k3,k4,k5,k6}
is the strongly anisotropic and large in magnitude heat conductivity 
of quasi-one dimensional compounds attributed to magnetic excitations. 
These experiments where partly motivated by the prospect of ballistic 
conductivity, as predicted for the $S=1/2$ Heisenberg model. Large 
thermal conductivities however were reported in ladder 
materials, e.g. $(La,Sr,Ca)_{14}Cu_{24}O_{41}$.
Note that the $S=1/2$ Heisenberg chain compounds offer the unique 
possibility to analyze 
directly the effect of spin-phonon scattering on thermal transport in a 
strongly interacting system as the spin-spin scattering is totally 
ineffective.

Thus, in connection to these experiments, we can conclude that: 

\begin{itemize}
\item They established a new, very efficient,  
mode of energy transport through magnetic excitations, similar 
in magnitude to metals but in insulators; the reason is that the 
characteristic magnetic exchange couplings are in the $eV$ range. 
\item These novel materials have potential for technological applications. 
\item They promote the study of prototype integrable models as the 
description of many of these compounds is in terms of, for example,  
the $S=1/2$ Heisenberg, Hubbard, sine-Gordon, nonlinear $\sigma$ model. 
\end{itemize}

\section{Perspectives}

From the above discussion we can conclude with a series of open issues 
and probable developments:

\begin{itemize}
\item The transport properties of integrable models are bound to 
be exactly evaluated following the development of appropriate Bethe ansatz 
techniques. These models offer the unique possibility of an 
analytical solution and it is ironical that because of their solvability 
they also present unconventional transport properties.

\item The transport of low energy effective theories, 
some of them integrable models, is to be settled.

\item The connection of high energy integrable models to their low energy 
effective field theories is to be clarified.

\item Of fundamental interest for the interpretation of experiments and 
applications is the role of perturbations in breaking the ideal transport.
Apparently unavoidable perturbations are phonons, disorder, 
interchain coupling. 

\item New experimental systems for the study of one dimensional quantum 
transport will be developed. For the time being magnetic compounds are the 
most promising, the electronic ones do not seem so ideal for fundamental 
studies.  But work is in progress in developing artificial one 
dimensional structures as superlattices or self-assembled 1D systems 
on surfaces that will offer far more controllable systems for the study 
of quantum transport.  
\end{itemize}

\section*{Acknowledgements}
On this occasion  I would like to thank the organizers of the 
Conference {\it Statistical Physics of Quantum Systems - novel orders and 
dynamics} held in Sendai in July 2004 for giving me the opportunity 
to present these ideas.
I would also like to thank my long standing collaborators in this field, 
P. Prelov\v sek, M. Long, J. Karadamoglou, H. Castella and F. Naef.

\end{document}